\documentclass[9pt]{IEEEtran}
\usepackage{cite}
\usepackage{graphicx}
\usepackage{subfig}
\usepackage{algpseudocode}
\usepackage{algorithm}
\usepackage{epstopdf}
\usepackage{color}
\usepackage{soul}
\usepackage{verbatim}
\usepackage{bbold}
\usepackage{balance}
\usepackage{kbordermatrix}
\usepackage{mathtools}

\makeatletter
\def\endthebibliography{%
  \def\@noitemerr{\@latex@warning{Empty `thebibliography' environment}}%
  \endlist
}
\makeatother

\begin{document}

\title{Weighted p-bits for FPGA implementation of probabilistic circuits }
\author{Ahmed Zeeshan Pervaiz, Brian M. Sutton, Lakshmi Anirudh Ghantasala, Kerem Y. Camsari}


\maketitle

\begin{abstract}
Probabilistic spin logic (PSL) is a recently proposed computing paradigm based on unstable stochastic units called probabilistic bits (p-bits) that can be correlated to form probabilistic circuits (p-circuits). These p-circuits can be used to solve problems of optimization, inference and also to implement precise Boolean functions in an ``inverted'' mode, where a given Boolean circuit can operate in reverse to find the input combinations that are consistent with a given output. In this paper we present a scalable FPGA implementation of such invertible p-circuits.  We implement a ``weighted'' p-bit that combines stochastic units with localized memory structures. We also present a generalized tile of weighted p-bits to which a large class of problems beyond invertible Boolean logic can be mapped, and how invertibility can be applied to interesting problems such as the NP-complete Subset Sum Problem by solving a small instance of this problem in hardware. 
\end{abstract}


\section{Introduction}\label{sec:Intro}

\IEEEPARstart{P}{robabilistic} spin logic (PSL) is a recently proposed computing paradigm based on unstable stochastic units called probabilistic bits (p-bit) that can be used to construct probabilistic circuits (p-circuits). These p-circuits can be used to solve a large class of problems including optimization \cite{sutton2017intrinsic}, inference \cite{behtash2016} and precise Boolean functions \cite{camsari2016,faria2017low,pervaiz2017hardware} to perform ``invertible'' logic. 


\par p-bits are  tunable random number generators (tunable RNG)  where a telegraphic output $m$ is controlled with an input $I$. For example a strong positive bias at the input will result in the output producing more highs than lows and vice versa. In mathematical terms, this behavior is described as:

\begin{equation}
\rm {m_i}(t) = {\rm{sgn}}\Big\{ \mathrm{rand(-1,1)} + \mathrm{tanh}\big({I_i}(t)\big)\Big\} 
\label{eq:p-bit}
\end{equation}
where rand($-$1,1) represents a random number from a uniform distribution between $-1$ to $1$. p-bits can be interconnected according to:

\begin{equation}
\rm {I_i}(t) = I_0\Big\{{h_i} + \sum_j {{J_{ij}}{m_j}(t)}\Big\} 
\label{eq:weight}
\end{equation}

\noindent where [J] is the interconnection matrix and $\rm \{h\}$ is the bias vector that adds a local contribution to each p-bit. $\rm I_0$ controls the strength of the interconnections that can function as an inverse pseudo-temperature of the system. (Eq.~\ref{eq:p-bit}) and (Eq.~\ref{eq:weight}) are the same as the defining equations for Boltzmann Machines introduced by Hinton and his collaborators \cite{ackley1985learning} which have had a tremendous impact on the field of machine learning.

\par A number of nanodevice implementations of p-bits (Eq.~\ref{eq:p-bit}) have been proposed using spintronic units such as stochastic magnetic tunnel junctions (MTJ) \cite{camsari2016,camsari2017edl}. Such stochastic MTJs have been experimentally demonstrated  \cite{grollier2016,majetich2016,mizrahi2015,fukushima2014spin,jp2014IEDM}. However, even though nanodevice implementations of p-bits could ultimately be more energy efficient and scalable, a large scale implementation of p-circuits based on nanodevices is difficult at present. Similarly, the interconnection matrix (Eq.~\ref{eq:weight}), like the p-bit, can be built with novel nanodevices such as crossbar arrays using memristors as envisioned in Ref.~\cite{pershin2010experimental} or with CMOS solutions such as those shown in Ref.~\cite{jarollahi2014nonvolatile,hu2015associative,pervaiz2017hardware}.  

\par In this paper we present a digital, tiled FPGA implementation of large p-circuits using a \textit{weighted} p-bit that combines the functionality of Eq.~\ref{eq:p-bit} and Eq.~\ref{eq:weight} in a single, composite unit where each weighted p-bit is a tunable random number generator that has a \textit{local} memory structure which weighs the outputs of other weighted p-bits. The main contribution of the present paper is to show how such weighted p-bits can be useful in tackling hard problems that are being explored in the context of alternative computational paradigms such as ``Ising''\cite{yamaoka201620k} and quantum computing. For example, integer factorization, which has been explored in the context of hardware quantum annealers\cite{dwave2016} using principles similar to ``invertible logic'' \cite{camsari2016}.


We further developed an $\sf n \times n$ array, demonstrated throughout with a specific 4$\times$ 4 example, on which one can map any  $\sf n^2 \times n^2$ [J] matrix. This 4$\rm \times $4 array can support a fully connected reciprocal network, a well known requirement for such networks like Boltzmann machines is the need to update the p-bits in a sequential manner \cite{hinton2007boltzmann}. To do so, we use a \textit{sequencer} that produces a set of enable signals for each p-bit in the 4$\rm \times $4 array. 

\par The tiles that we use to build larger p-circuits are generally reciprocal networks \cite{hinton2007boltzmann,hinton2006fast,hinton2006reducing} that resemble the architecture of Ising machines \cite{yamaoka201620k,yoshimura2016fpga,ortega2016fpga,wang2017oscillator,mcmahon2016fully,inagaki2016coherent, sutton2017intrinsic,belletti2008simulating,belletti2009janus}, however in general p-circuits can be constructed as a \textit{directed} network of reciprocal subcircuits \cite{camsari2016} that are not Ising machines, as we describe in detail in Section \ref{sec:res}.


\par The organization of this paper is as follows: In Section~\ref{sec:wp-bit} we describe the FPGA implementation of the weighted p-bit. In Section~\ref{sec:res} we demonstrate examples of p-circuits realizing invertible Boolean logic starting from simple Boolean gates  that are then interconnected to construct an N-bit  invertible Ripple Carry Adder in Section~\ref{sec:Nbit} and a small instance solver for the NP-complete Subset Sum Problem in Section~\ref{sec:SubsetSum}.  

\section{Weighted p-bit} \label{sec:wp-bit}

\par Fig.~\ref{fig:pbit} shows the block diagram of an FPGA implementation of a weighted p-bit. There are two major sub-blocks of the weighted p-bit: (a) The weight matrix which implements Eq.~\ref{eq:weight} and (b) the tunable RNG which implements Eq.~\ref{eq:p-bit}.  We describe both components below. 

\subsection{Weight Matrix}

\par Each weighted p-bit can take the outputs of others and weight them according to the interconnection ([J]) matrix. This is done by every weighted p-bit locally using a weight matrix block that takes the $i^{th}$ row of the [J] matrix and stores it in registers local to the weighted p-bit along with the $i^{th}$ entry of the bias vector, $\rm \{h\}$. The local presence of these registers allows a compact implementation of Eq.~\ref{eq:p-bit}-\ref{eq:weight} in a single unit. These registers can also be made user accessible, however in our demonstration this functionality is not needed since all our [J] and $\rm \{h\}$ entries are obtained offline without any online learning \cite{Schuman2017ASO}. Moreover, the examples discussed  in this paper do not  make use of ``annealing'' \cite{sutton2017intrinsic}, which would also require user accessibility.


\par  \textbf{Fixed point arithmetic:} To perform all arithmetic operations in a weighted p-bit,  we use a fixed point notation of s[\textit{x}][2], where integer $x$ is chosen based on the requirements of the p-circuit. This allows a range of $-|2^x|$ to $(2^x-1)$ for the integer part. For example, the [J] and $\{h\}$ matrices used for the Full Adder shown in Fig.~\ref{fig:FullAdder} use weights that require s[4][2] while for an AND gate shown in Fig.~\ref{fig:AND_floating}, the use of s[3][2] is sufficient.  In general,  [J] and $\rm \{h\}$ registers allow different problems to be mapped onto the system and having a wide range of allowed weights enables a broader category of problems to be solved.

\par
\begin{figure}[t]
\centering
{\includegraphics[width=\linewidth]{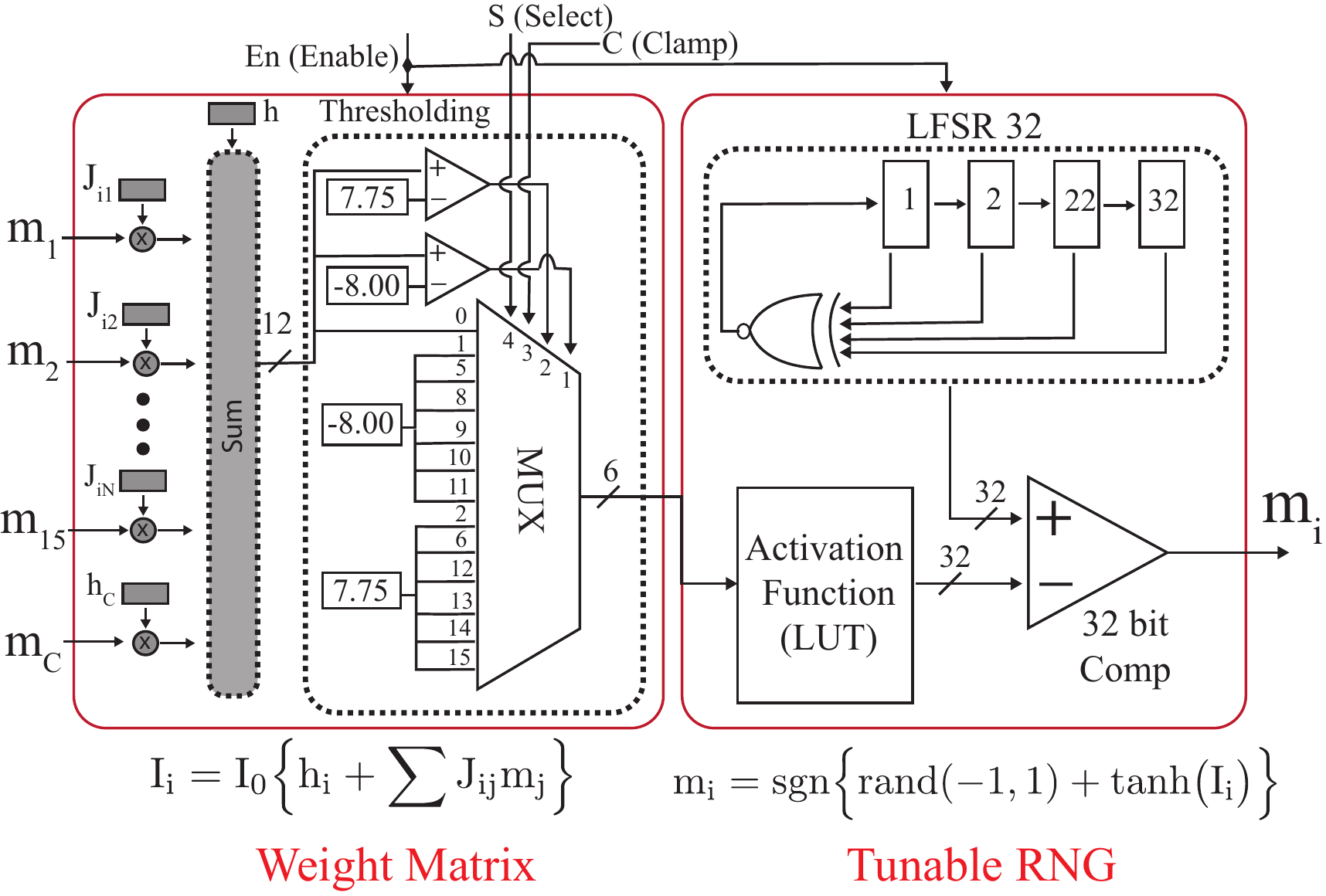}}
\caption{\textbf{Weighted p-bit:} A weighted p-bit consists of two major subblocks, a \textit{Weight Matrix} and a \textit{Tunable RNG}  implementing Eq.~\ref{eq:weight} and Eq.~\ref{eq:p-bit} as a composite unit. The weight matrix implements one column of Eq.~\ref{eq:weight} and adds overflow protection and clamping capabilities to the weighted p-bit while the tunable RNG subblock implements Eq.~\ref{eq:p-bit} whose terminal characteristics are further shown in Fig. \ref{fig:Sigmoid}. See text for a detailed description.}
\label{fig:pbit}
\end{figure}

 \textbf{Thresholding:} Given that each weighted p-bit has multiple inputs, the worst case for the weighted sum $\rm I_0([J] m + \{h\})$ can exceed the allowed input range of s[x][2] notation or the allowed input range of the \textit{Activation Function}, (which uses the output of the weighted sum block to calculate tanh as explained in the subsequent subsection.) To prevent this, an overflow detection and numerical clamping system are used that compare a bit extended result from the \textit{Sum} block to the maximum and minimum allowed numbers for the \textit{Activation Function}.  The result of the sum is clamped to the the maximum or the minimum number that can be read by the \textit{Activation Function}. For example, the s[4][2] notation has a maximum and minimum limit of 15.75 and -16, however the input of the lookup table for the \textit{Activation Function} need not be any less than -8 and any greater than 7.75 as shown in (Fig.~\ref{fig:pbit}). 

\par \textbf{MUX:} A multiplexer is used to perform both the thresholding and clamping of weighted p-bits (Fig.~\ref{fig:pbit}). Table~\ref{table:Mux} shows the truth table of the multiplexer. Four signals are used as inputs, ``S(Select)'', which is high if the weighted p-bit is to be clamped to the ``C(Clamp)'' signal. The other two are the outputs of signed comparison between the Sum and the maximum/minimum numbers to be passed on to the \textit{Activation Function}.

\begin{table}[b]
\setlength{\tabcolsep}{3pt}
\renewcommand{\arraystretch}{1.3}
\centering
\begin{tabular}{||c| c| c| c| c||}
\hline
 S (Select)  &  (C) Clamp  &   { $\rm I_{IN}> max_{tanh}$}  &    {$\rm I_{IN}< min_{tanh}$} &  Output     \\
 
(4) & (3) & (2) & (1) & \\

 \hline
0 & x & 0 & 0 & $\rm I_{IN}$\\

 \hline
0 & x & 0 & 1 & $\rm min_{tanh}$\\

 \hline
0 & x & 1 & 0 & $\rm max_{tanh}$ \\

 \hline
1 & 0 & x & x & $\rm min_{tanh}$ \\

 \hline
1 & 1 & x & x & $\rm max_{tanh}$ \\

 \hline \hline
\end{tabular}
\caption{ Truth Table for the weight matrix multiplexer}
\label{table:Mux}
\end{table}

\subsection{Tunable random number generator}  
  
\par
\begin{figure}[t]
\centering
{\includegraphics[width=0.80\linewidth]{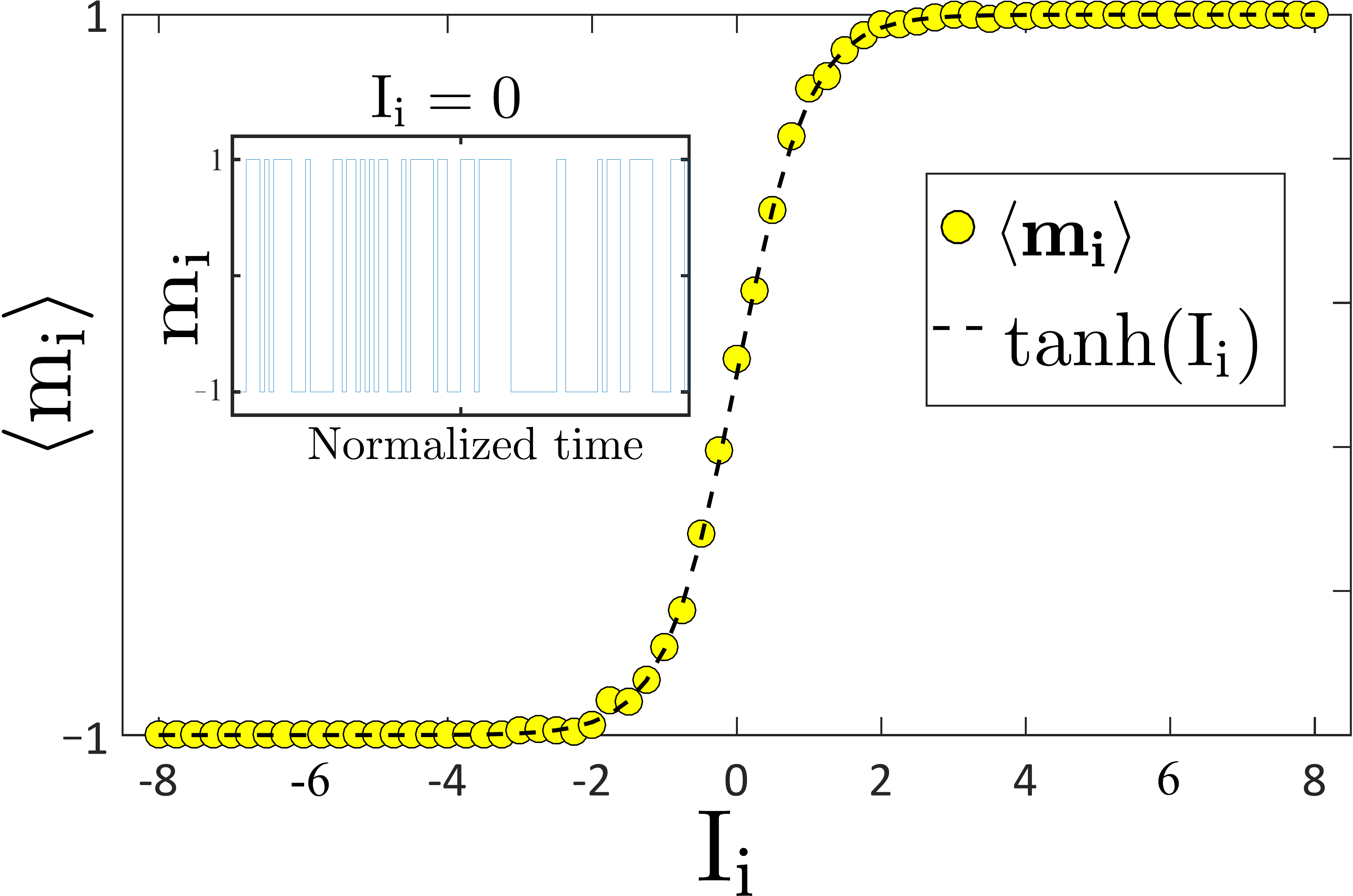}}
\caption{\textbf{Sigmoid:} The time-averaged output ($\rm m_i$) of a weighted p-bit  is shown as a function of the applied input $\rm I_i$. When the $\rm I_i=0$ (inset), the output $\rm m_i$ shows equal amounts of 1's and 0's with a long-time average of 0.5. As $\rm I_i$ is increased above (below) 0, the average increases and saturates to 1 ($-$1). Here, the binary output of the FPGA $m_i \in \{0,1\}$ is converted to a bipolar $\rm m_i \in \{-1,+1\}$ representation. }
\label{fig:Sigmoid}
\end{figure}

 The output of the weight matrix is applied as input to the tunable RNG. Fig.~\ref{fig:Sigmoid} shows the average time characteristics of the tunable RNG block. As shown in the inset of Fig.~\ref{fig:Sigmoid}, when the input $\rm I_i$ is 0, the $m_i$ randomly fluctuates between 0 and 1 as a function of time,  leading to a long time average of 0.5. As the applied input is increased above (below) 0, the average increases (decreases) and saturates to 1 (-1). The tunable randomness allows  weighted p-bits to become correlated with each other. We describe the submodules of the tunable RNG block below.

\par \textbf{Activation Function:} We use lookup tables (LUT) to implement the {tanh} function. The domain of the tanh is $ (-\infty,+\infty)$ and its range is $\rm (-1,1)$. To allow a comparison between the output of the pseudo random number generator and that of the LUT, we first transform tanh to $ z=(\mathrm{tanh}+1)/2$ and then use a s[0][31] bit fixed-point representation, where 0 represents the integer part and 31 represents the fractional part. We choose a s[3][2] representation for the input of the LUT, which translates to the interval $(-8,7.75)$ with a resolution of 0.25 between successive data points. Several methods of implementing sigmoid like functions have been studied\cite{tommiska2003efficient} and lookup tables  are naturally suited for implementation in FPGAs.The use of lookup tables will result in an approximation error. Ref. \cite{tommiska2003efficient} looks at the average and the maximum errors for various approximation methods. For the lookup table an error arises due to the difference in absolute real number value of tanh and its truncated representation which is shown in Ref. \cite{tommiska2003efficient} (Eq. 9) as $E_{trunmax}=2^{-(b+1)}$ for a fixed point representation of s[a][b].

\par \textbf{Pseudo-random number generation:} We use a 32-bit Linear feedback shift register (LFSR) with an XNOR feedback to the first register using taps from 32, 22, 2 and 1 position in the LFSR \cite{xilinx_doc}. Given a seed value, this produces a maximal length pseudo-random stream of size $ 2^{32} - 1$ with the all 1's  being the only state that is not part of the stream. It is important to note that each weighted p-bit in a p-circuit must have a unique seed value, otherwise the p-bits may have unintentional strong correlations resulting in incorrect system operation. In this paper, the use of more complex pseudo-RNGs were avoided due to the complexity of implementation and size. In practice, LFSR based pseudo-RNG worked well and are naturally suited for digital implementation.

\par \textbf{Comparator:} A 32-bit comparator compares the outputs of the \textit{Activation Function} and the pseudo random number generator and  produces 0 or 1 state at the output, as shown in the inset of Fig.~\ref{fig:Sigmoid}.  
\vspace{-11pt}

\subsection{System Tile}
 \par
\begin{figure}[t]
\centering
{\includegraphics[width=0.90\linewidth]{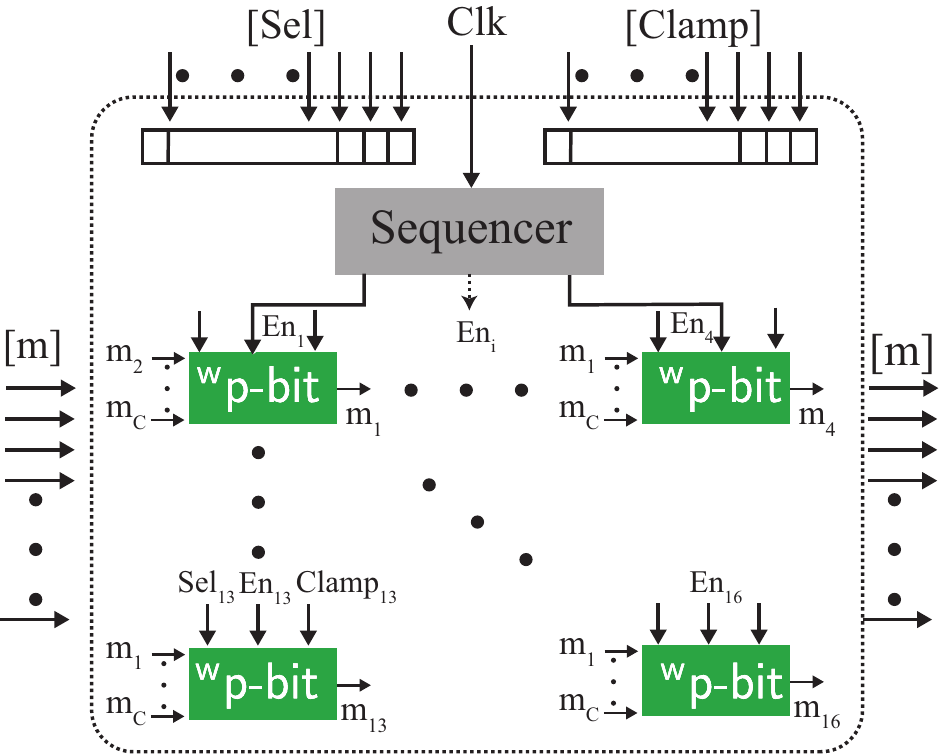}}
\caption{\textbf{4 x 4 system tile:} A 4 $\times$ 4 block of weighted p-bits (denoted by $\sf ^wp$-bit) can be used to implement [J] matrices with a dimension of $\rm 4^2 \times 4^2$ and the sequencer block allows each of the 16 p-bits to be updated sequentially for proper system operation. Different problems can be mapped through a choice of suitable [J] and $\rm \{h\}$ matrices to construct larger p-circuits.}
\label{fig:Tile}
\end{figure}

 \textbf{Serial updating:} Our p-circuits within a tile are similar to reciprocal networks and in general reciprocal networks, such as unrestricted Boltzmann machines, require all p-bits to be updated sequentially\cite{hinton2007boltzmann}. To ensure this requirement is met, a ``sequencer'' is present in each p-circuit which generates an \textit{Enable} signal for every p-bit in the p-circuit, ensuring no two p-bits are active simultaneously. At any given point in time, only one of these enable signals is high while all others are kept low, allowing only one p-bit to update. An AND gate (as implemented here) has 3 p-bits with each p-bit requiring 2 clock cycles for a complete update. To help ensure timing closure within the FPGA, a gap of 1 clock cycle is present between adjacent \textit{Enable} signals. In this case the update order is ($\rm A \rightarrow B \rightarrow C)$ but this sequence itself could have been randomized at  each iteration as a method to avoid unwanted correlations due to the update order. In general, the updating sequence could also influence the average time it takes for the p-circuit to settle to its steady-state, but this is not discussed further. A specific example which illustrates the connectivity within a tile ( such as the 4 $\times$ 4 shown in Fig. \ref{fig:Tile}) is shown in Fig. \ref{fig:ANDGate}, with all the connections presented.

\par \textbf{Mapping problems} Another important aspect of the system tile is to allow mapping of different problems\cite{vinci2015quantum}\cite{choi2011minor}. In this manuscript we demonstrate simple boolean gates such as an AND gate and a Full Adder which require 3 and 5 (or 14) p-bits for functioning. For every problem there is a necessary requirement of serial updating which is fulfilled using the sequencer present within a tile, but larger problems such as the 32-bit Ripple Carry Adder can be implemented by cascading system tiles that implement smaller p-circuits. This cascading in a parallel manner results in a serial-parallel architecture. The presence of this serial-parallel architecture preserves the invertibility of the boolean gates all the while allowing a speed up as the problems are scaled. The ideal size of a system tile is dictated by the maximum network size of coupled p-bits which require serial updating. Note that if the network size is larger than the tile size, multiple tiles can be joined using minor graph embedding\cite{vinci2015quantum}\cite{choi2011minor}, though with some trade-offs. For example a p-circuit that requires 5 p-bits with p-bits being updated serially, needs only a system tile of size 5, not any larger. Many such 5 p-bit systems tiles can be put together to build a larger instance of the same problem. However, note that it is also possible that some problems would require all p-bits to be updated serially, in which case the entire problem can only be mapped onto a single system tile. Note that in such a case the solution will inherently be slower in a synchronous implementation, since the time for a complete update of a system tile increases linearly with the number of p-bits. For example, the AND gate presented in the section III-A needs $\rm 3 \times (2+1)$ clock cycles for a complete update, while the Full Adder requires $\rm 5/14 \times (2+1)$ clock cycles (since we present a 5 and a 14 p-bit Full Adder design) for a complete update. On the other hand, the 32-bit Ripple Carry Adder presented in the section III-C requires the same $\rm (5,14) \times (2+1)$ clock cycles (depending on which Full Adder design is used) for an update since it is  using 32 individual Full Adders connected in parallel. In general, the size of the system tile and the speed of one complete update  depend on the details of how the problem is mapped.

\par \textbf{Interconnecting System tiles} For certain problems, scaling to larger instances will be possible by interconnecting system tiles. The Ripple Carry Adder and the Subset Sum solver implemented in this manuscript are two such examples where Full Adders realized within a system tile are interconnected to form larger more complex systems. In such problems, the system tiles need to be connected in a ``directed'' manner where the strength of the connection can be manipulated. For example in the 32-bit RCA, the Carry out of a Full Adder is connected to the Carry In of the proceeding Full Adder. This connection can be done via the following

\begin{enumerate}
\item The Select and Clamp signals shown in Fig. 1. For example for the 32-bit Adder one could Clamp the Select line of all Carry-in p-bits and clamp them to the Carry-out line from the preceding Adder with the exception of the First Full Adder which has its Carry-in clamped to 0. 
\item By the \textbf{$\rm m_C$} terminal. This terminal allows the output of a p-bit to be weighted by an interconnect strength $\rm h_C$ such that when $\rm h_C \rightarrow \infty$ the p-bit is effectively cloned to the signal coming into $\rm m_C$, while for $\rm h_C \rightarrow 0$, the signal $\rm m_C$ has no effect on the  p-bit operation. Note that $\rm h_C \rightarrow \infty$ is the same as using the Select and Clamp signal as presented previously.
\end{enumerate}
\subsection{FPGA}
\par {\textbf{I/O Architecture for FPGA:} We use the Xilinx Kintex ultrascale  XCKU040-1FBVA676 FPGA. The Xilinx Vivado Design Suite was used to synthesize and implement the Verilog RTL for the FPGA. As shown in Fig.~\ref{fig:System}, I/O operations with the p-circuits was accomplished by memory-mapping the p-circuits using AXI peripheral logic. Once wrapped, a number of standard interfaces can be used to control and extract data from the FPGA. Herein, we used a standard UART connection coupled to a Xilinx MicroBlaze softcore processor. For simplicity, we targeted a base operating frequency of 100 MHz for the design, as the principle objective was to explore invertible logic using p-circuits. For high-performance directed Boolean logic, we believe an optimized CMOS design would be more appropriate, however, since there is no equivalent of invertible Boolean logic in CMOS, we believe that the real application space for p-circuits lies in this domain. Table~\ref{table:Resource} presents a summary of resource utilization of the various designs that have been implemented in this paper. 
\begin{figure}[t!]
\centering
{\includegraphics[width=0.85\linewidth]{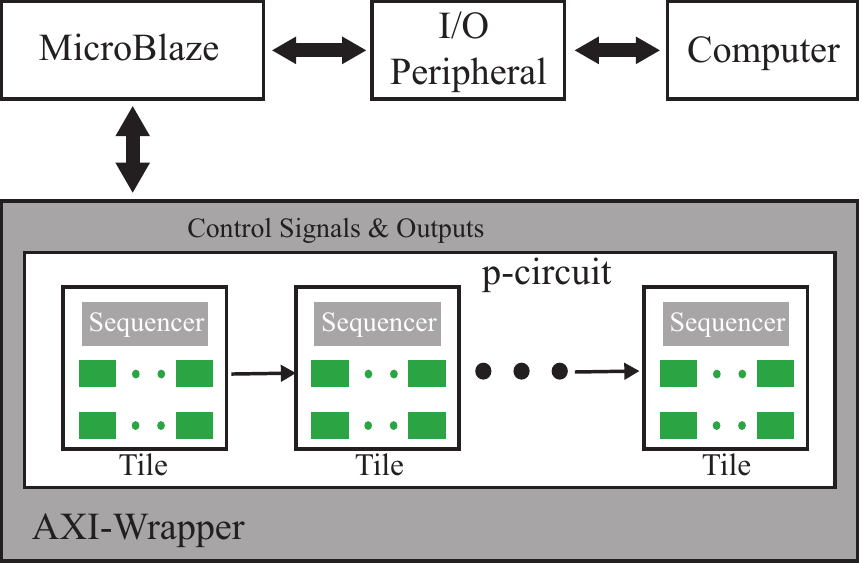}}
\caption{\textbf{I/O Architecture for FPGA:} We communicate with the input and output terminals of weighted p-bits using an I/O architecture whose block diagram is shown above. Any p-circuit (tile or collection of tiles) can be converted into an AXI (universal serial bus architecture) peripheral, which can then communicate with a computer via a MicroBlaze processor that allows the collection of data from p-circuits.}
\label{fig:System}
\end{figure}

\begin{table}[b]
\renewcommand{\arraystretch}{1.3}
\centering
\begin{tabular}{||c| c| c| c||}
\hline
\bfseries  &  \bfseries  Total  &   \bfseries Slice  &   \bfseries Slice     \\
  &  \bfseries  Weighted p-bits &  \bfseries  LUTs &  \bfseries Registers    \\
 
 \hline
\textbf{Kintex Ultrascale} & & 242400 & 484800   \\ 
XCKU040-1FBVA676 & &  &     \\ 

\hline
\textbf{Tunable RNG} & 1 & 42 & 33 \\
\hline
\textbf{AND Gate} & 3 & 156 & 123   \\

\hline
\textbf{Full Adder} & 14 & 1345 & 586   \\

\hline
\textbf{15-bit SSP problem} & 155 & 14931 & 7083   \\

\hline
\textbf{32-bit Ripple Carry Adder} & 434 & 38814 & 18071  \\

\hline \hline
\end{tabular}
\caption{FPGA resource utilization of the p-circuits that have been implemented in this paper. }
\label{table:Resource}
\end{table} 

\section{Results}
\label{sec:res}

\subsection{AND Gate}\label{sec:AND_Gate}

\begin{figure}[t]

\centering
\subfloat[]{\includegraphics[width=0.99\linewidth]{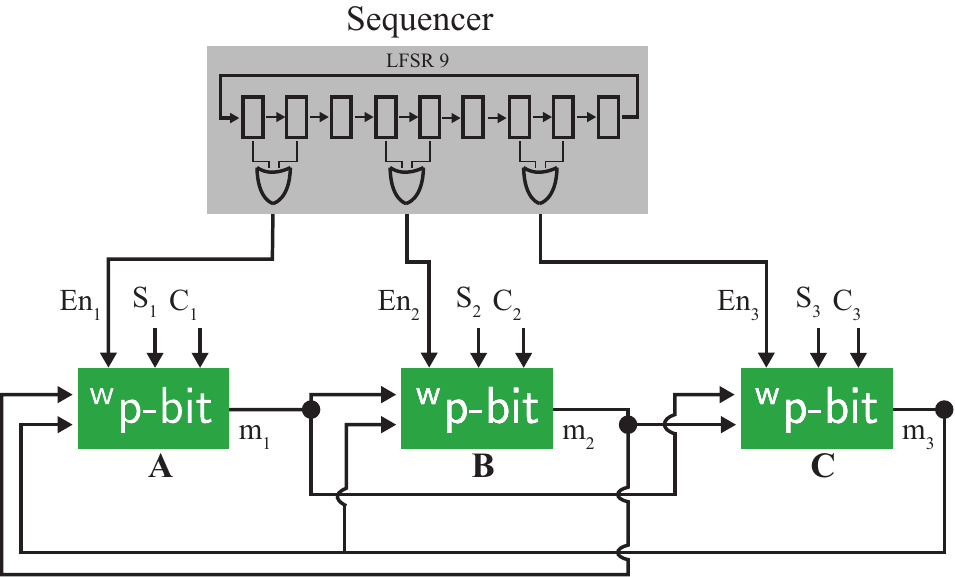}}

\caption{\textbf{AND Gate:} Three weighted p-bits are needed to implement an AND gate whose [J] matrix is obtained from Ref. \cite{biamonte2008nonperturbative}. A sequencer circuit is used to force an updating  sequence of ($\rm A \rightarrow B \rightarrow C)$.}
\label{fig:ANDGate}
\end{figure}

\par Fig.\ref{fig:ANDGate} shows the block diagram of an AND gate that is implemented using 3 p-bits with each p-bit having two inputs.  The weighting matrices [J] and $\{h\}$ are from Ref. \cite{biamonte2008nonperturbative} and shown below: 

\renewcommand{\kbldelim}{(}
\renewcommand{\kbrdelim}{)}
\begin{equation} 
J_{AND} = 
  \kbordermatrix{
  & A & B & C \\
  & 0 & -1 & 2\\
  &-1 & 0 & 2\\
  & 2 & 2 & 0  
  }
,\hspace{.5cm} h^T = 
\begin{pmatrix}
1 & 1 & -2
\end{pmatrix}
\end{equation}

The p-circuit architecture of Ref. \cite{camsari2016} forces m to be  bipolar, i.e. m $\in \{1,-1\}$. It is more convenient to work with  a binary representation of  1 and 0, i. e m~$\in \{0,1\}$, in the FPGA which requires that the [J] and $\rm \{h\}$ matrices be mapped to binary bases. This can be accomplished by the following transformation: $\rm J_{\rm binary}=2 \times J_{\rm bipolar}$ and $\rm h_{binary}=h_{bipolar} - J_{bipolar}\mathbb{1} $, where $\mathbb{1}$ is an all ones vector of size $N\times 1$.

\begin{figure}[t]
\centering
\includegraphics[width=0.99\linewidth]{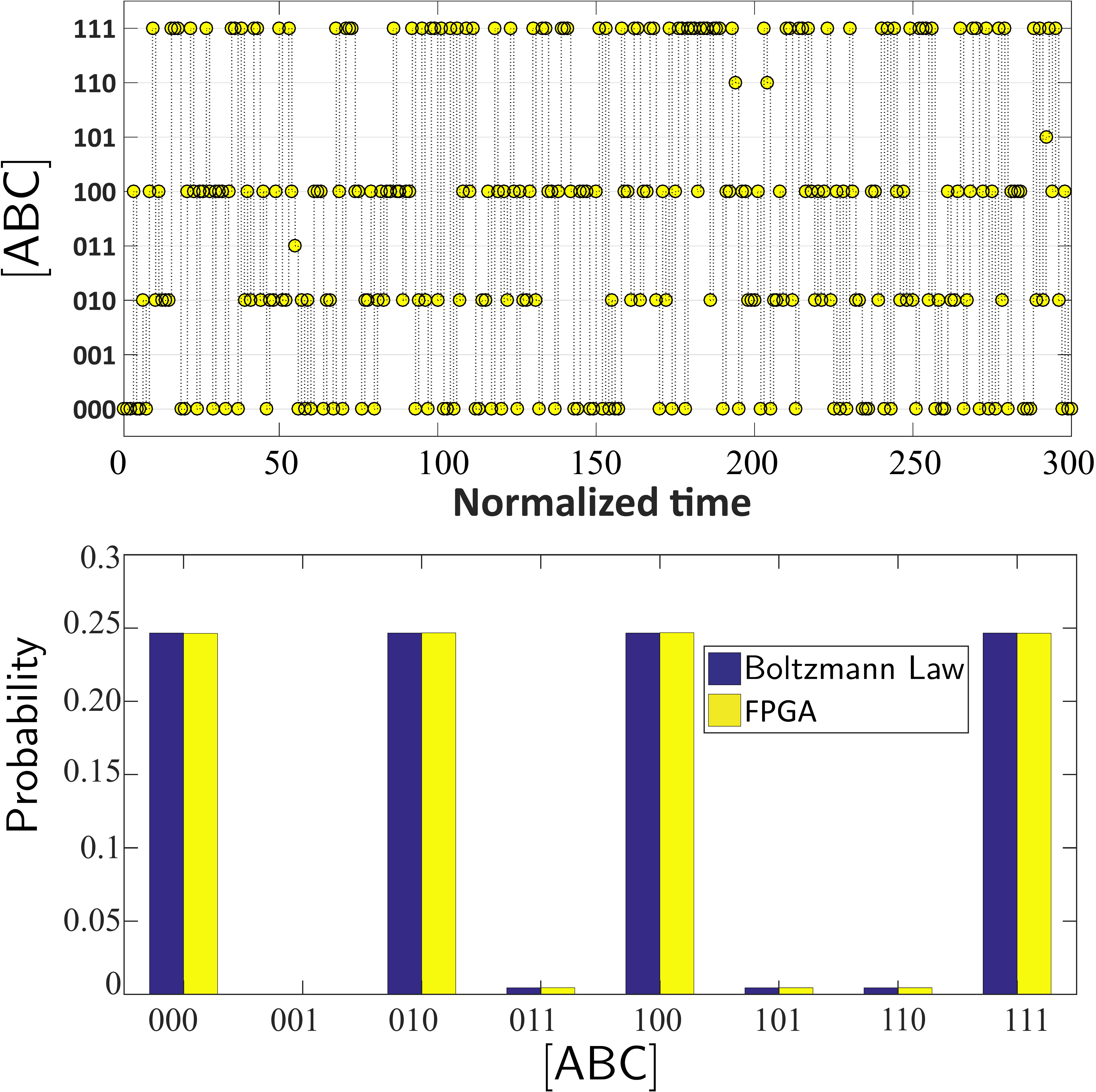}
\caption{\textbf{Floating mode (AND):} (a) Time dependent outputs of [ABC] for the  AND gate are shown as a function of samples collected from the serial port of the FPGA. (b) The weighted p-bits are correlated and when left floating they reproduce the truth table of the AND Gate as shown by the time-averaged statistics which are collected using $\rm 10^6$ samples. The FPGA results are in excellent agreement with the Boltzmann Law of Eq.~\ref{eq:BL}.}
\label{fig:AND_floating}
\end{figure}

\par \textbf{Floating mode (AND):} Fig.~\ref{fig:AND_floating} shows the operation of an AND gate with all weighted p-bits left floating, where the states $\rm [ABC]$ corresponding to the truth table ($\rm A \cap B = C)$ of an AND gate are visited with high probability. Note that this is a unique property of p-circuits with no counterpart in a digital CMOS implementation of an AND gate. In reciprocal networks with symmetric [J] matrices, an energy functional E for the state $\rm \{m\}=[m_i,m_j,\cdots] ^T $ can be defined as \cite{hinton2007boltzmann, camsari2016}:

\begin{equation}
\mathrm{
E(\{m\}) =- I_0 \bigg(\sum_{i,j} \frac{1}{2} \left(J_{ij} m_i m_j\right) + \sum_i h_{i} m_i\bigg)}
\end{equation} 

Then, Boltzmann Law describes the steady state probabilities for each configuration $\rm\{m\}$ according to,

\begin{equation}
\mathrm{
P(\{m\}) = \frac{\exp(-E(\{m\}))}{ \sum_{i,j} \exp(-E(\{m\}))}}
\label{eq:BL}
\end{equation}

\par Fig. \ref{fig:AND_floating} shows the steady state statistics of the AND gate in excellent agreement with the Boltzmann Law for a total of $\rm 10^6$ samples.

\par \textbf{Forward / Invertible mode (AND):} Fig. \ref{fig:AND_clamped}(a) shows the statistics for the system when both inputs A and B have been clamped to 1 through the Select and Clamp signals that control the bias vector $\{h\}$. In this case, for the chosen $\rm I_0$,  the output C mostly stays high (1) which means the circuit is operating like a standard digital AND gate. A remarkable property of p-circuits is in their input/output equivalence similar to the gates discussed in the context of memcomputing \cite{traversa2017}. The output bits (C) can also be clamped and in this case the inputs (A,B) fluctuate among  combinations consistent with the clamped output. Fig.~\ref{fig:AND_clamped}(b) shows the long time statistics of the system when the output C has been clamped to 0. It can be seen that the system spends an equal amount of time visiting three possible combinations of (A,B), namely (0,0), (0,1) and (1,0). This basic example can be imagined to be  1-bit factorization  of an AND gate where the factors of the product 0 are identified. 

\begin{figure}[t]
\centering
{\includegraphics[width=0.85\linewidth]{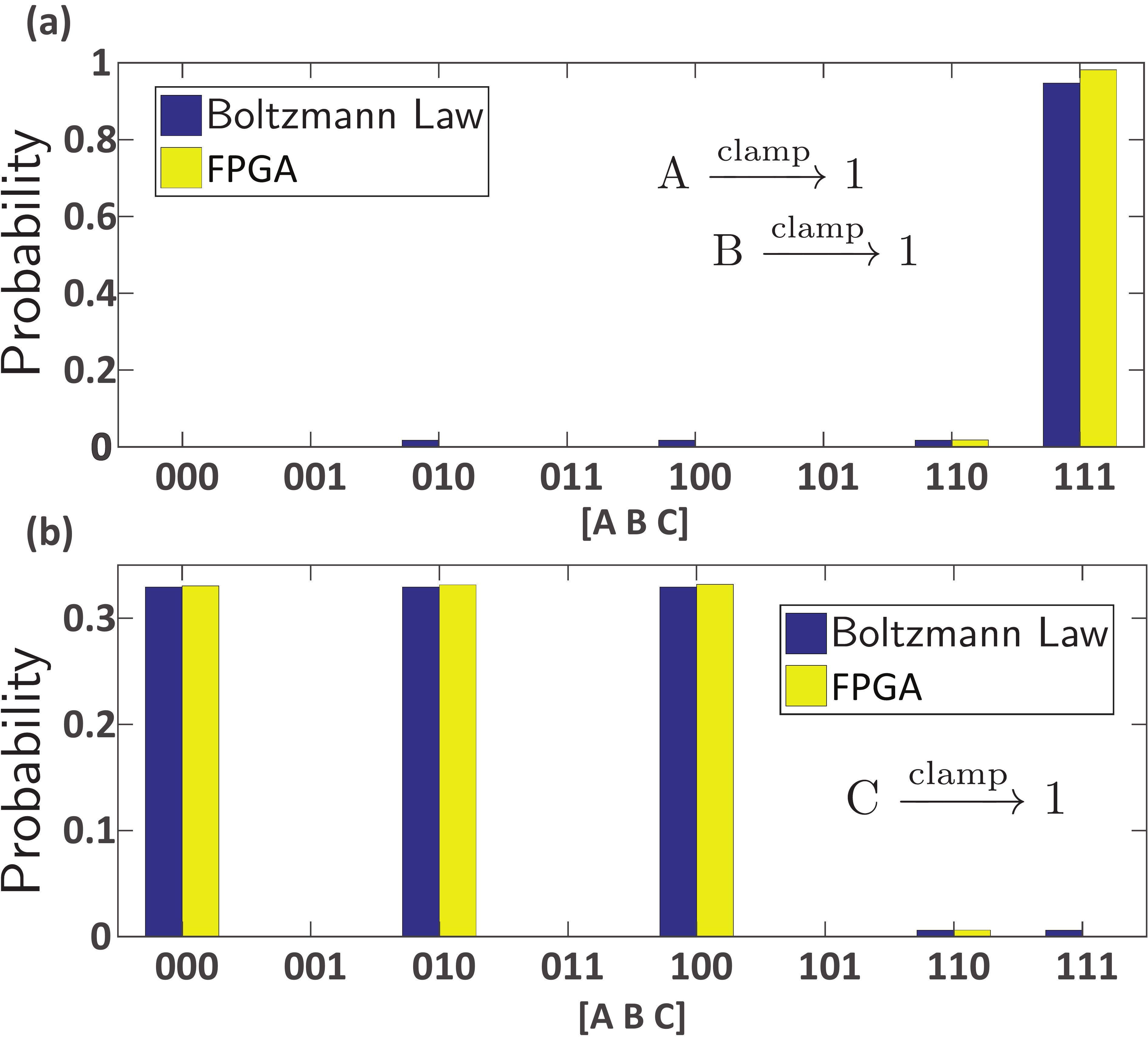}}
\caption{\textbf{Forward / Invertible mode (AND):} Any weighted p-bit of the AND gate, input (A, B) or output (C) can be clamped using the bias vector $\rm \{h\}$ through the Select and Clamp signals. (a) shows the long time statistics when the inputs A and B have been clamped to 1, while (b) shows the long time statistics when the output C has been clamped to 0. In both cases $\rm 10^6$ samples have been used.  }
\label{fig:AND_clamped}
\end{figure}

\subsection{Full Adder}\label{sec:Full_Adder}

\par A Full Adder was implemented as a p-circuit following the architecture of Ref. \cite{camsari2016}. In Ref.~\cite{camsari2016} 14 p-bits are used to build a Full Adder, of which only 5  constitute input/output terminals, namely $ \rm C_{IN}, A, B, Sum$ and $\rm C_{OUT}$. The remaining 9  are known as  ``auxiliary'' p-bits. In this paper, we improve the 14 p-bit implementation of the invertible Full Adder (FA) in Ref. \cite{camsari2016} and implement the same functionality using 5 p-bits. This is achieved by first noting that the first half of the truth table is complementary to the second half for the FA (Fig 8). The first 4 lines in the truth table are then turned into an orthonormal set by a Gram-Schmidt process and a [J] matrix is obtained using Eq. 12 in Ref. \cite{camsari2016} which is finally rounded to integer values, with diagonal entries replaced by zeroes
\renewcommand{\kbldelim}{(}
\renewcommand{\kbrdelim}{)}
\begin{equation} 
J_{FA} = 
  \kbordermatrix{
    & C_{in} & B & A & S & C_{out} \\
   & 0 & -1 & -1 & 1 & 2  \\
   &-1 & 0 & -1 & 1 & 2 \\
   &-1 & -1 & 0 & 1 & 2  \\
   &1 & 1 & 1 & 0 & -2 \\
   &2 & 2 & 2 & -2 & 0 
  }
\end{equation}

These designs for the Full Adder fit within the 4$\times$4 tiles that were defined previously with less packing efficiency, but since the design is reconfigurable, appropriate changes can be made at a relatively low cost by scaling the tile size up or down. Similar to the AND gate we convert the weight matrix into their binary equivalents using the transformation shown earlier. A summary of resource utilization for the 14 p-bit Full adder is given in Table~\ref{table:Resource}.

\par Fig.~\ref{fig:FullAdder} shows the state of a 14 p-bit Full Adder when all the p-bits have been left floating. The truth table of the Full Adder is highlighted in floating mode and can be seen by the statistics shown in Fig.~\ref{fig:FullAdder} that are collected using $\rm 10^6$ samples, once again in excellent agreement with the Boltzmann Law. Due to its sequential updating, this Full Adder design requires $ \rm 14 \times [2 + 1 (gap)] = 42$ clock cycles for one complete update. The Full Adder is the largest p-circuit that we have built within a 4$\times$4 tile, since each weighted p-bit within the Full Adder needs to be updated sequentially. In the next section we use this 14 p-bit Full Adder to construct a N-bit Ripple Carry Adder, while in section III-D we use a 5-bit Full Adder to solve a small instance of the SSP.  

\begin{figure}[t]

\centering
{\includegraphics[width=0.99\linewidth]{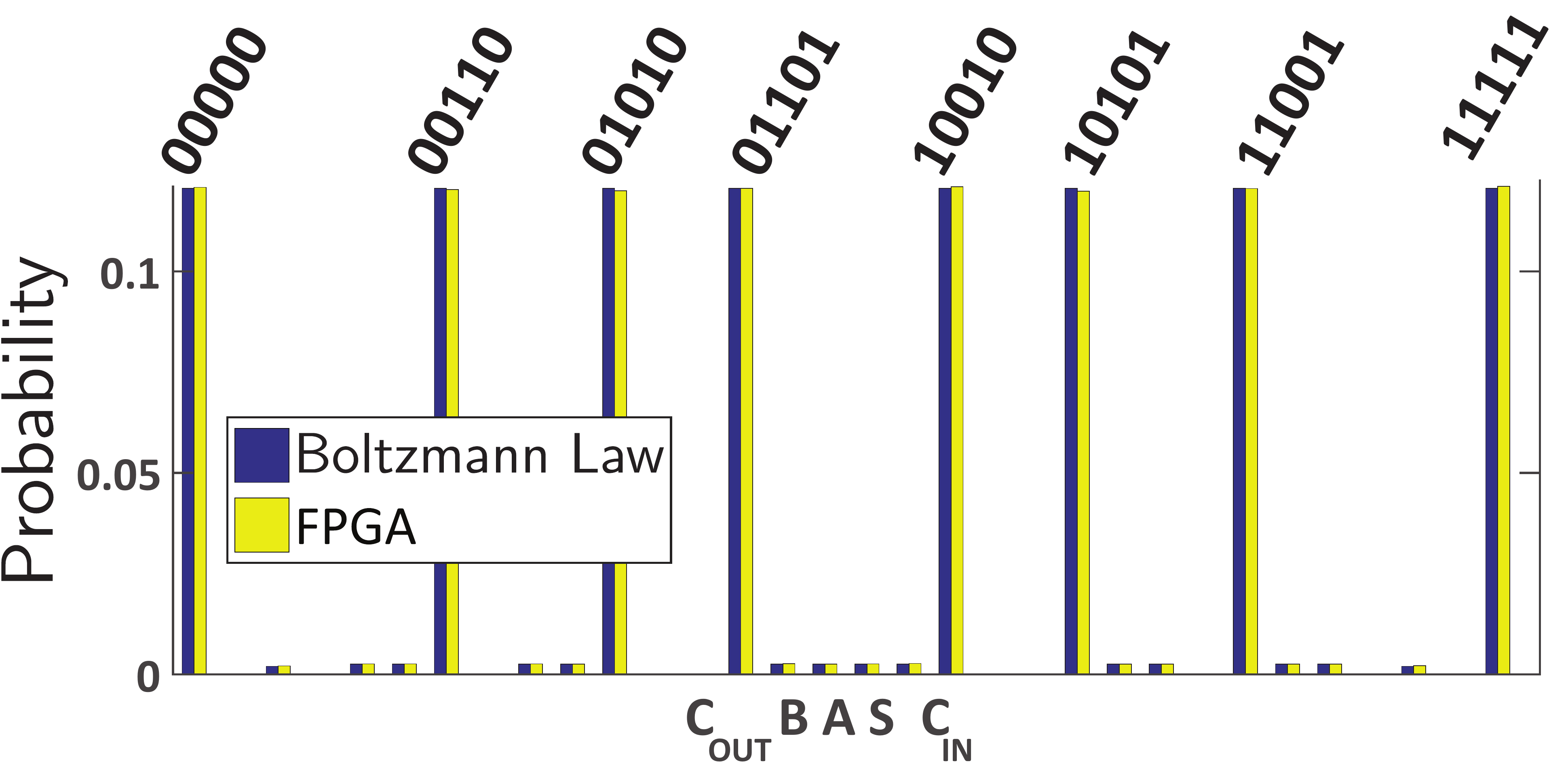}}
\caption{\textbf{Floating mode (Full Adder):} Long time statistics of the 14 weighted p-bit Full Adder using $\rm 10^6$ samples when all the terminals have been left floating are shown in the figure above. Similar to the AND gate, the Full Adder reproduces its truth table when all p-bits are left floating. Results show excellent agreement with Boltzmann Law (Eq.~\ref{eq:BL}).}
\label{fig:FullAdder}
\end{figure}

\subsection{N-bit Ripple Carry Adders}\label{sec:Nbit}
Unlike the reciprocal networks ($\rm J_{ij}=J_{ji})$ we have shown so far, we now construct a \textit{directed} p-circuit by cascading the symmetric Full Adders in a \textit{parallel} architecture without any global sequencer circuit.
This is very different from the AND gate and Full Adder presented in sections III-A and III-B which are designed within a  $4\times 4$ tile, where each p-bit is updated sequentially. This serial-parallel update scheme significantly speeds up convergence time. 

Fig. \ref{fi:fig9}a shows the block diagram of multiple tiles, each designed as a Full Adder, that are interconnected in a directed way to form an N-bit Ripple Carry Adder (RCA). What makes the RCA directed is the fact that the carry out bit of the Full Adders is connected only from the least significant bit to the most significant bit  but not vice versa. Note that this constitutes a significant difference from the AND and Full Adder  because the Boltzmann Law is not applicable to this system anymore.


\par  In this design we chose a fully directed connection between Full Adders. However, in general, tiles can be interconnected via an adjustable connection that can be partially (or fully) bidirectional using the terminal $m_c$ shown in Fig. \ref{fig:pbit}. This concept of directionality is key to building larger p-circuits as shown in Ref. \cite{camsari2016} where a strong degree of bidirectionality can lead to erroneous results unless a very large number of time samples are obtained. 

\par The system shown Fig.~\ref{fi:fig9}a in general produces a sum (S) consistent when the inputs A, B are clamped to an N-bit number functioning as an adder. 
However, the system can also function as a subtractor when an N-bit input (A or B) and the sum are clamped to a given number, even though the system is no longer completely bidirectional. Fig.~\ref{fi:fig9}b shows the long time statistics for the N-bit Ripple Carry Adder when all N-bit terminals (S=sum, A and B) have been left floating and, remarkably, the system correlates in such a manner to select  a single state (S$-$A$-$B=0) with $\approx 20\%$ probability out of $10^5$ samples. This feature can be used to solve hard problems such as the 3-sum problem that is concerned with finding a set of inputs (A, B, C) that add up to a given sum S \cite{cormen2009}. With minor modifications, the invertible Full Adders could also be used to solve the Subset Sum Problem, using similar adder architectures shown in \cite{traversa2017}. A digital implementation such as our invertible N-bit adder could be used to solve such problems in hardware very efficiently.

\par The N-bit Ripple Carry Adder presented in Fig.~\ref{fi:fig9} is \textit{not} a true sequentially updated machine because while each adder takes 42 clock cycles to produce one complete update, the adders themselves do not wait $\rm 42 \times N$ for one complete update. In this way the N-bit Adder presented in Fig.~\ref{fi:fig9}  is a serial-parallel architecture different from the serially updated N-bit adders presented in Ref.~\cite{camsari2016}. This serial-parallel update for the N-bit RCA seems to operate accurately for the deterministic update sequence we chose, but it is not clear if this approach would be generally applicable for any problem, which is beyond the scope of this paper. While we do not give a quantitative analysis of the speed up from this serial-parallel architecture, we note that in the case of the invertible N-bit RCA, this serial-parallel architecture combined with fast clock speeds of the FPGA should allow considerably faster operation of this large scale p-circuit as compared to computer simulations.  

\begin{figure}[t]
\centering
{\includegraphics[width=0.95\linewidth]{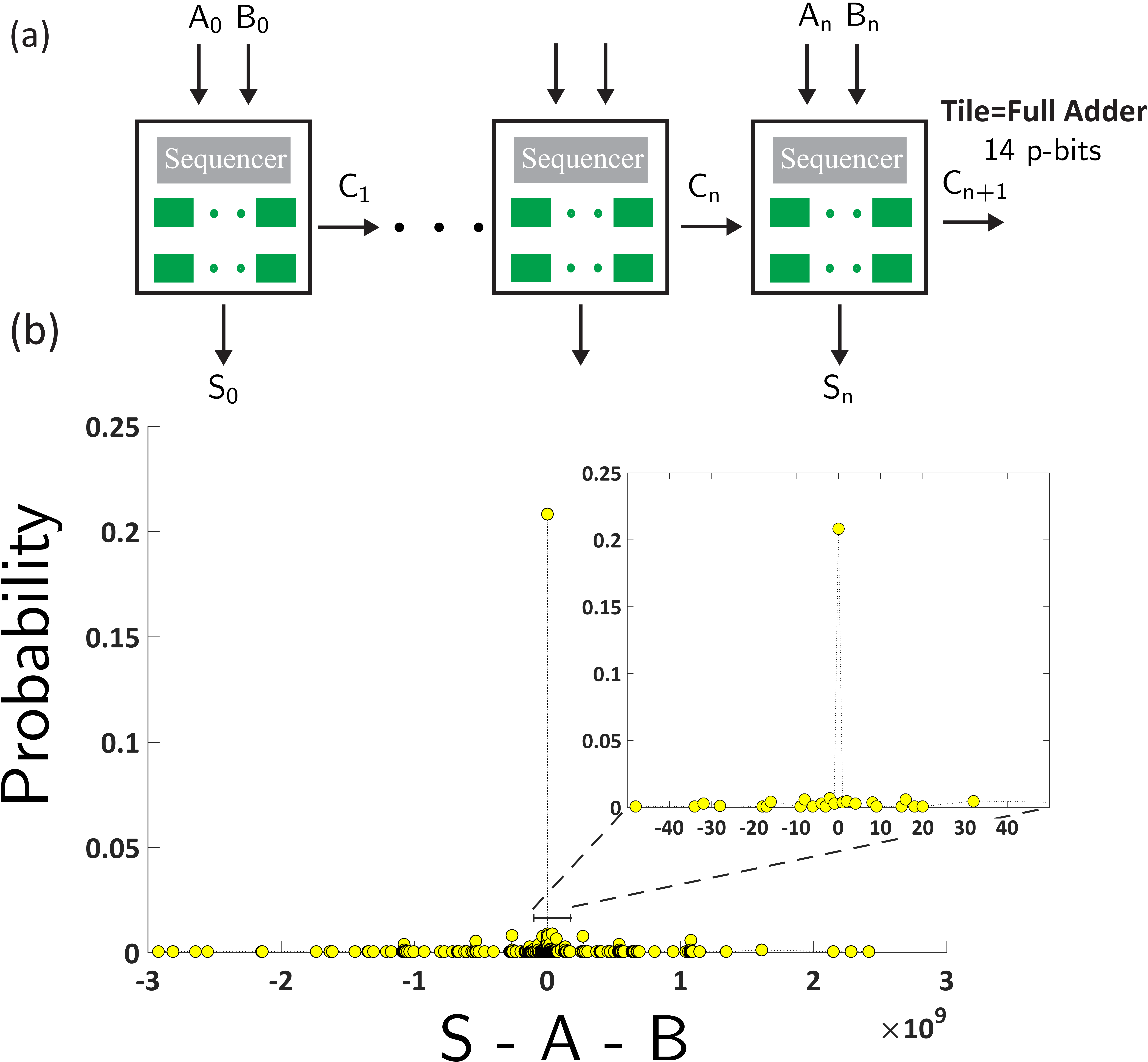}}
\caption{\textbf{N-bit Ripple Carry Adder:} (a) Reciprocal networks of individual tiles programmed as Full Adders are interconnected in a directed manner to construct an N-bit Ripple Carry Adder (RCA). (b) The RCA is left floating and the long time statistics of the N-bit sum (S) and the inputs (A, B) get correlated in such a way to make a \textit{single state} (inset) S$-$A$-$B = 0 appear with $\approx$ 20\% probability out of $10^5$ samples among billions of states ($\pm 2^{32})$, as can be seen in the $x$-axis. Only $\approx 1500$ samples are shown for clarity. } 
\label{fi:fig9}
\end{figure}

\begin{figure}[t]
\centering
{\includegraphics[width=0.99\linewidth]{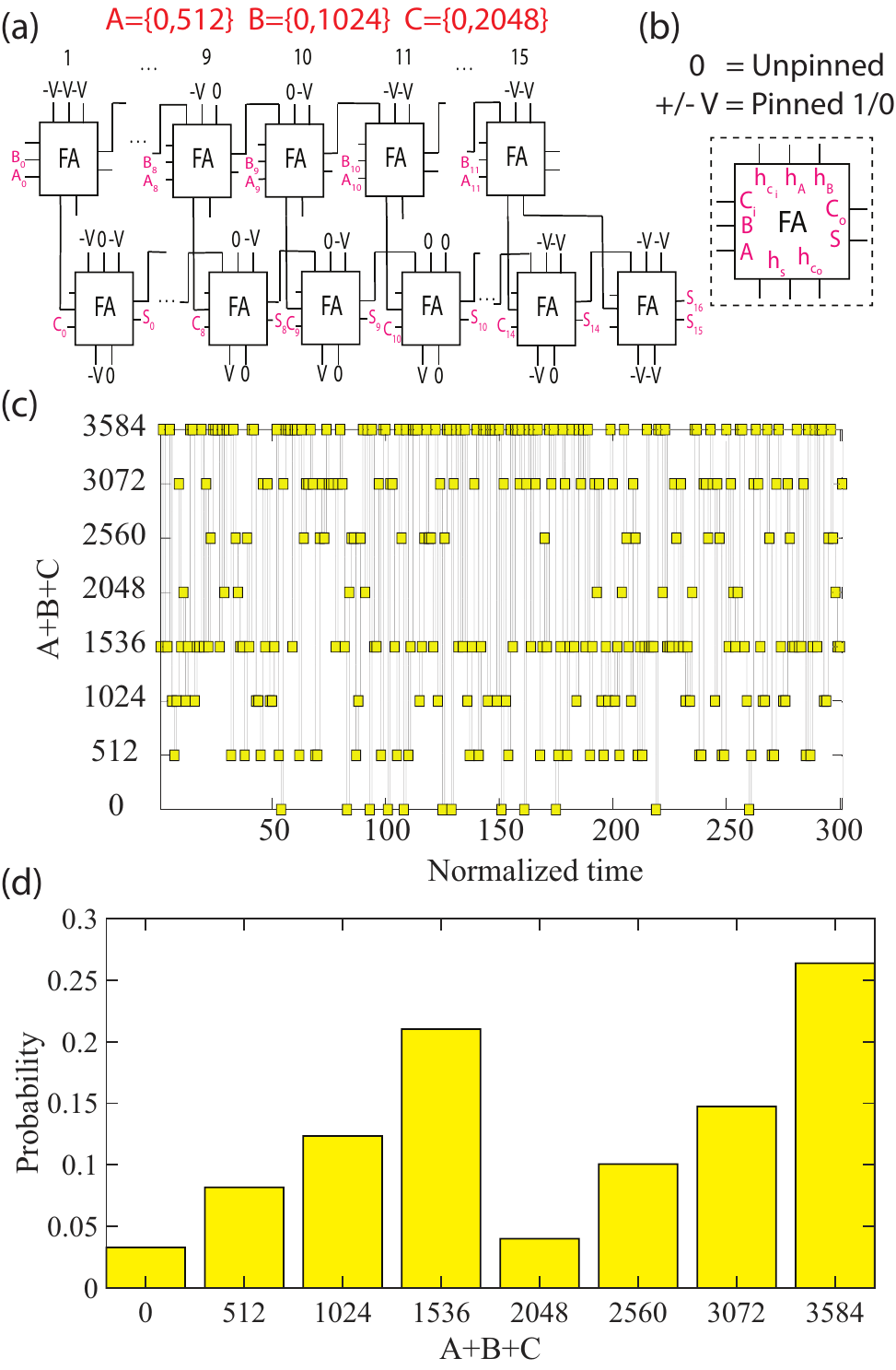}}
\caption{\textbf{Subset Sum Problem:} (a) An adder that adds three 15-bit numbers A,B and C to give a 17-bit Sum S. The Sum S is first clamped to a particular number which in this case is 3584. The inputs A, B and C are constrained to particular sets using a scheme shown in (b) for each bit of the inputs. Note how the connections from the bottom layer of adders are \textit{directed} where the sum is clamped to the top layer where the inputs A and B are added. In this example A is $\{0,512\}$, B is $\{0,1024\}$, and C is $\{0,2048\}$. (c) shows 300 time samples taken from a data sequence of $10^5$ from which two values of 3584 and 1536 appear more then the other 6 possible states. (d) shows the histogram corresponding to $10^5$ samples.} 
\label{fi:fig10}
\end{figure}

\subsection{Subset Sum Problem}\label{sec:SubsetSum}

\par In this subsection we show how the Full Adder block and the serial-parallel architecture of the N-bit Ripple Carry Adder can be used to solve a small instance of the NP-complete Subset Sum Problem (SSP) \cite{cormen2009}. In this problem, a set G with a finite number of positive numbers is defined, and from this set the problem is to determine whether there exists a subset $\rm S'$ such that $\rm S' \subseteq G$ has elements which sum to a specific target S. Figure.~\ref{fi:fig10} shows a circuit that can be programmed to select a 17-bit sum S, while the 15-bit inputs are constrained to particular sets. In the example shown in Fig.\ref{fi:fig10}, the sum S is set to 3584 while the inputs A, B and C are constrained to the sets  $\{0,512\}$, $\{0,512\}$ and  $\{0,512\}$ respectively. Note that in Fig.~\ref{fi:fig10}(b) we show terminals $h_{XX}$ for ease of visualization where $h=-V$ means that the \textit{select} and \textit{clamp} lines are connected to 1 and 0 respectively, while $h=+V$ means that the \textit{select} and \textit{clamp} lines are connected to 1 and 1 respectively, and $h=0$ means that the select line is connected to 0. One striking feature of this circuit is that the flow of information in the structure shown in Fig.~\ref{fi:fig10}(a) is upwards, i.e. information flows from the Sum to inputs A and B. 

\par The inputs can be constrained to sets by clamping  certain bits to 0 or 1 depending on the choice of the set. For the input A used in Fig.~\ref{fi:fig10}, all the bits except the $9^{th}$ from the LSB side are clamped to 0. Clamping the bits hence allows the inputs to be \textit{constrained} to particular sets, forcing the circuit to look within a certain configuration consistent with the members of the set. Fig.~\ref{fi:fig10}(c) shows 300 time samples of A+B+C taken from a data sequence of $10^5$ when the Sum is clamped to 3584. In this case where the correct inputs are  A=512, B=1024, and C=2048, it can be seen that two states one correct 3584 and another incorrect 1536 appear closer to each other and far removed from the other 6 possible combinations that the system could be in. The relative probabilities of different peaks in the solution (such as 1536 and 3584) is a function of the inverse pseudo-temperature ($I_0$ in Eq.~4) and could be made larger by increasing this value. However, in practice this could cause the system to get stuck in a meta-stable state for a long time. Therefore, for this example we have chosen a relatively small $I_0=1$ that does not      create a large difference in probabilities. Fig.~\ref{fi:fig10}(d) shows the statistics for the entire $10^5$ samples from which the state 3584 has a higher peak than 1536. We note that this particular example of the SSP is easily solvable and does not constitute a hard instance. Our main purpose is to illustrate how invertible Full Adders can be interconnected to design a hardware solver for this problem, similar in spirit to the approach described in Ref.~\cite{traversa2017}. 

\section{Conclusion}
We have presented a digital tiled FPGA implementation of probabilistic circuits with which we demonstrate examples of invertible Boolean logic. We have used a weighted p-bit design where neuron and synapse-like functionalities are combined in a single composite unit. The digital tiled nature of our architecture will allow p-circuits to scale as the FPGA densities scale, and we show how these tiles can be combined to construct large p-circuits such as the N-bit Ripple Carry Adder and a solver for the NP-complete Subset Sum Problem which use the invertibility of p-circuits. 

\section{Acknowledgment}
\footnotesize The authors gratefully acknowledge useful discussions with Supriyo Datta. This work was supported in part by C-SPIN, one of six centers of STARnet, a Semiconductor Research Corporation program, sponsored by MARCO and DARPA.

\bibliographystyle{IEEEtran}
\balance


\end{document}